\documentstyle[twoside,fleqn,espcrc2,epsbox]{article}
\newcommand{\AmS}{{\protect\the\textfont2
  A\kern-.1667em\lower.5ex\hbox{M}\kern-.125emS}}
\title{Duality in Landau-Zener-Stueckelberg potential curve crossing}
\author{Kazuo Fujikawa%
\address{Department of Physics, University of Tokyo,\\
Bunkyo-ku, Tokyo 113, Japan}%
\thanks{Invited talk presented at 7th Asia Pacific Physics
Conference, August 19-23, 1997, Beijing, China (to be published in
the Proceedings)}
and Hiroshi Suzuki%
\address{Department of Physics, Ibaraki University,\\
Mito 310, Japan}}
\begin{document}
\begin{abstract}
It is pointed out that there exists an interesting strong and weak
duality in the Landau-Zener-Stueckelberg potential curve crossing. A
reliable perturbation theory can thus be formulated in the both
limits of weak and strong interactions. It is shown that main
characteristics of the potential crossing phenomena such as the
Landau-Zener formula including its numerical coefficient are
well-described by simple (time-independent) perturbation theory
without referring to Stokes phenomena. A kink-like topological object
appears in the ``magnetic'' picture, which is responsible for the
absence of the coupling constant in the prefactor of the Landau-Zener
formula. It is also shown that quantum coherence in a double well
potential is generally suppressed by the effect of potential curve
crossing, which is analogous to the effect of Ohmic dissipation on
quantum coherence.
\end{abstract}
\maketitle
%%%%%%%%%%%%%%%%%%%%%%%%%%%%%%%%%%%%%%%%%%%%%%%%%%%%%%%%%%%%%%%%%%%%
\section{Introduction}
\label{sec:one}

The potential curve crossing is related to a wide range of physical
and chemical processes, and the celebrated Landau-Zener
formula~\cite{1,2,3} correctly describes the qualitative features of
those processes~\cite{4,5,6,7,8,9,10}. It has been recently
shown~\cite{11} that the potential crossing problem contains
interesting modern field theoretical ideas, namely, the duality and
gauge transformation.

The adiabatic and diabatic pictures in potential curve crossing
problem are related to each other by a field dependent $su(2)$~gauge
transformation~\cite{5,8}, and we point out that this transformation
leads to an interchange of strong and weak potential curve crossing
interactions, which is analogous to the electric and magnetic duality
in conventional gauge theory~\cite{12}. This strong and weak duality
allows a reliable perturbative treatment of potential curve crossing
phenomena at the both limits of very weak (adiabatic picture) and
very strong (diabatic picture) potential crossing interactions.

%%%%%%%%%%%%%%%%%%%%%%%%%%%%%%%%%%%%%%%%%%%%%%%%%%%%%%%%%%%%%%%%%%%%
\section{A model Hamiltonian of potential curve crossing and duality}
\label{sec:two}

\begin{figure}[htb]
\vspace{9pt}
%\framebox[55mm]{\rule[-21mm]{0mm}{43mm}}
\psbox[width=60mm]{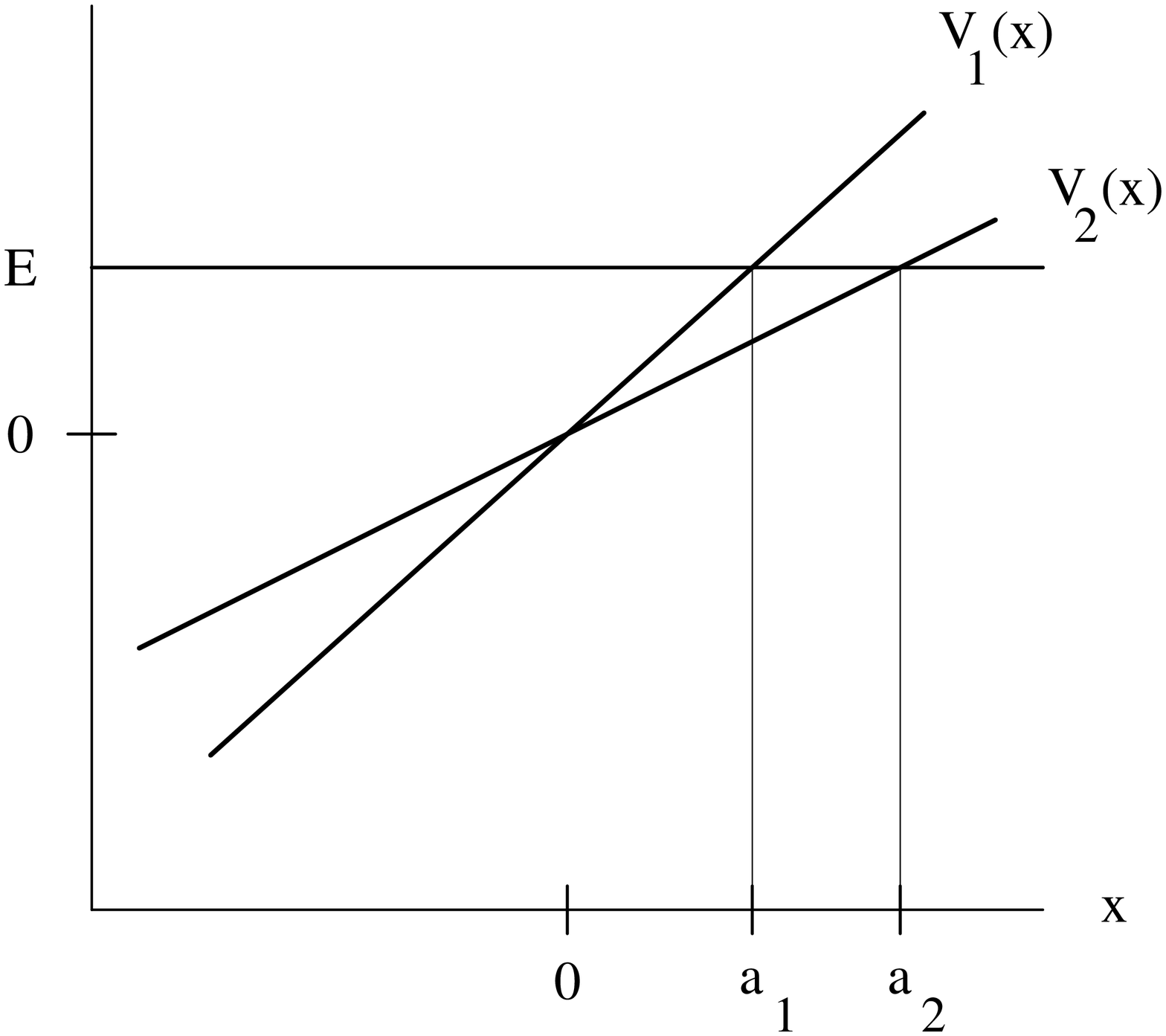}
\caption{Landau-Zener process in the diabatic (electric) picture.}
\label{fig:largenenough}
\end{figure}
\begin{figure}[htb]
%\framebox[55mm]{\rule[-21mm]{0mm}{43mm}}
\psbox[width=60mm]{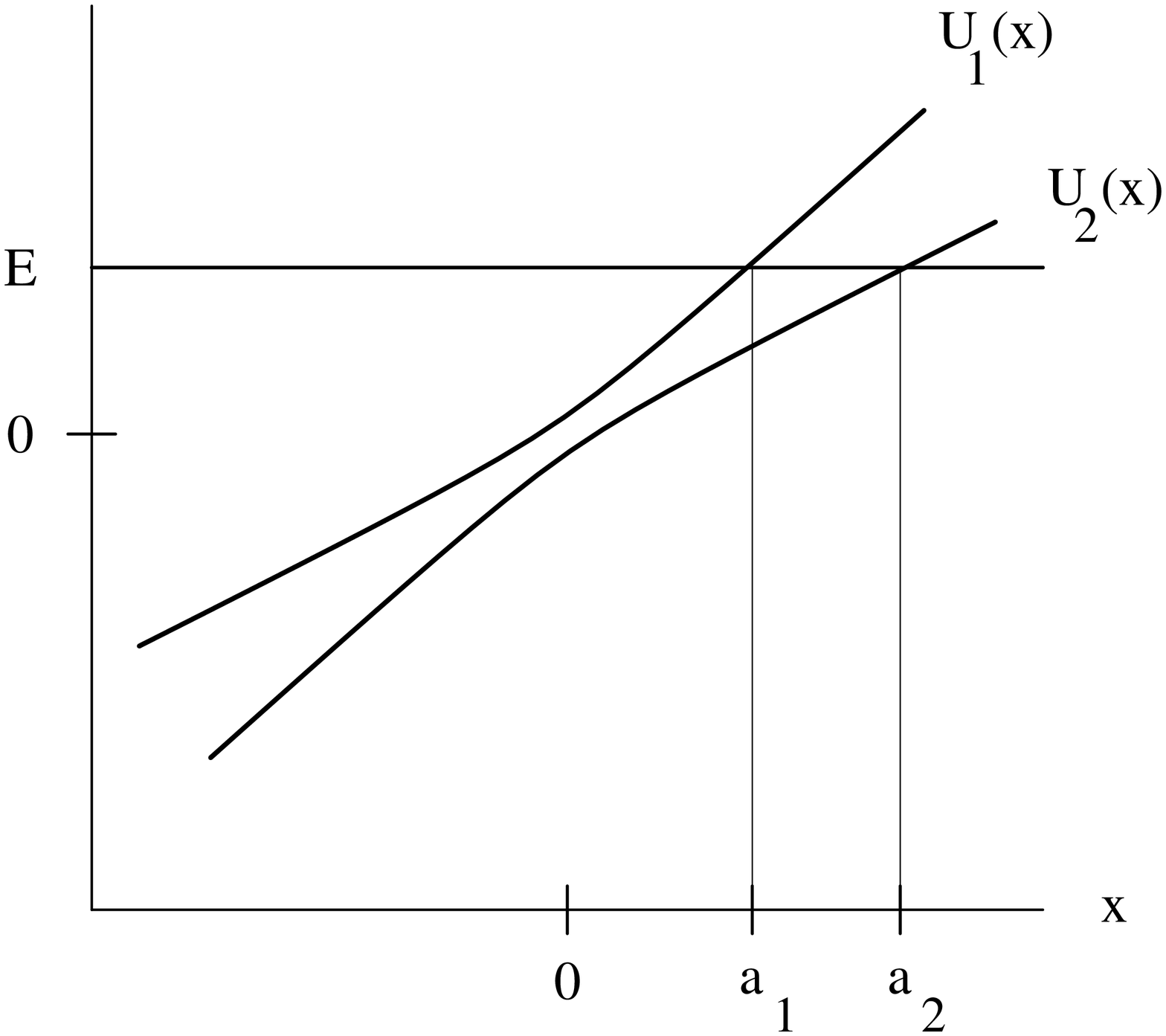}
\caption{Landau-Zener process in the adiabatic (magnetic) picture.}
\label{fig:toosmall}
\end{figure}

To analyze the potential curve crossing, we start with a model
Hamiltonian defined in the so-called diabatic picture~\cite{5,8}
\begin{eqnarray}
   H&=&{1\over2m}\hat p^2+{V_1(x)+V_2(x)\over2}
   +{V_1(x)-V_2(x)\over2}\sigma_3
\nonumber\\
   &+&{1\over g}\sigma_1
\label{eq:two.one}
\end{eqnarray}
where $\sigma_3$ and~$\sigma_1$ stand for the Pauli matrices. We
assume throughout this article that the potential crossing occurs at
the origin, $V_1(0)=V_2(0)=0$ (see~Fig.~1).

If one neglects the last term in the above Hamiltonian, one obtains
the unperturbed Hamiltonian in the diabatic picture
\begin{equation}
   H_0\equiv{1\over 2m}\hat p^2+{V_1(x)+V_2(x)\over2}
   +{V_1(x)-V_2(x)\over2}\sigma_3.
\label{eq:two.two}
\end{equation}
This Hamiltonian~$H_0$ describes two potentials, which are decoupled
{}from each other. The last term in~(\ref{eq:two.one}),
$H_I\equiv\sigma_1/g$ with a constant~$g$, causes the transition
between these two otherwise independent potential curves. In other
words, if one takes $g\rightarrow{\rm large}$, this case physically
corresponds to a {\it complete\/} potential crossing from a view
point of {\it adiabatic\/} two-potential crossing in~Fig.~2. Namely,
$g$~stands for the strength of potential crossing interaction, and
$g\rightarrow{\rm large}$ corresponds to a very strong potential
crossing interaction. On the other hand, if one lets $g$~smaller, the
effects of the last term in~(\ref{eq:two.one}) become substantial and
the Hamiltonian $H_0$~(\ref{eq:two.two}) does not present a sensible
zeroth order Hamiltonian.

To deal with the case of a small~$g$, we perform the non-Abelian
``gauge transformation,''
\begin{eqnarray}
   \Phi(x)&=&e^{i\theta(x)\sigma_2/2}\Psi(x),
\nonumber\\
   H'&=&e^{i\theta(x)\sigma_2/2}He^{-i\theta(x)\sigma_2/2},
\label{eq:two.three}
\end{eqnarray}
where $\sigma_2$~is a Pauli matrix. The Hamiltonian in the new
picture is given by
\begin{eqnarray}
   H'&=&{1\over2m}
   \left[\hat p-{\hbar\over2}\partial_x\theta(x)\sigma_2\right]^2
   +{V_1(x)+V_2(x)\over2}
\nonumber\\
   &+&\left[{V_1(x)-V_2(x)\over2}\cos\theta(x)
                          +{1\over g}\sin\theta(x)\right]\sigma_3
\\
   &+&\left[-{V_1(x)-V_2(x)\over2}\sin\theta(x)
                          +{1\over g}\cos\theta(x)\right]\sigma_1.
\nonumber
\label{eq:two.four}
\end{eqnarray}
To eliminate the potential curve mixing, the last term
of~(\ref{eq:two.four}), we choose the gauge parameter~$\theta(x)$
as~\cite{8}
\begin{equation}
   \cot\theta(x)=g{V_1(x)-V_2(x)\over2}\equiv f(x).
\label{eq:two.five}
\end{equation}
We then obtain the Hamiltonian in the {\it adiabatic\/} picture
\begin{equation}
   H'=H_0'+H_I',
\label{eq:two.six}
\end{equation}
where
\begin{eqnarray}
   H_0'&\equiv&{1\over2m}\hat p^2+{U_1(x)+U_2(x)\over2}
\nonumber\\
   &+&{U_1(x)-U_2(x)\over2}\sigma_3,
\label{eq:two.seven}
\end{eqnarray}
and
\begin{eqnarray}
   H_I'&\equiv&-{\hbar\over4m}
   \left[\hat p\partial_x\theta(x)
         +\partial_x\theta(x)\hat p\right]\sigma_2
\nonumber\\
   &+&{\hbar^2\over8m}\left[\partial_x\theta(x)\right]^2.
\label{eq:two.eight}
\end{eqnarray}
The potential energies in the adiabatic picture are related to those
in the diabatic picture as (Fig.~2)
\begin{eqnarray}
   U_{1,2}(x)&\equiv&{V_1(x)+V_2(x)\over2}
\nonumber\\
   &\pm&\sqrt{\left[{V_1(x)-V_2(x)\over2}\right]^2+{1\over g^2}}.
\label{eq:two.nine}
\end{eqnarray}
{}From the definition of the gauge parameter in~(\ref{eq:two.five}),
the ``gauge field''~$\partial_x\theta(x)$ is expressed as
\begin{equation}
   \partial_x\theta(x)=-{f'(x)\over1+f(x)^2}.
\label{eq:two.ten}
\end{equation}

The transition from the diabatic picture to the adiabatic picture is
a field dependent transformation.

In the adiabatic picture, the $\sigma_2$~dependent term in the
interaction~$H_I'$~(\ref{eq:two.eight}) causes the potential 
crossing. If one neglects~$H_I'$, the two potentials characterized by
$U_1(x)$ and~$U_2(x)$ do not mix with each other: Physically, this
means {\it no\/} potential crossing. This suggests that $H_I'$~is
proportional to the coupling constant~$g$, since a small~$g$
corresponds to {\it weak\/} potential crossing by definition. This is
in fact the case as is clear from (\ref{eq:two.ten})
and~(\ref{eq:two.five}).

We thus conclude that the two extreme limits of potential crossing
interaction should be reliably handled in perturbation theory;
namely, the strong potential crossing interaction in the
{\it diabatic\/} picture, and the weak potential crossing interaction
in the gauge transformed {\it adiabatic\/} picture. This is analogous
to the electric-magnetic duality in conventional gauge
theory~\cite{12}: The diabatic picture may correspond to the electric
picture with a coupling constant~$e=1/g$, and the adiabatic picture
to the magnetic picture with a coupling constant~$g$.

A general criterion for the validity of perturbation theory in the
adiabatic picture~(\ref{eq:two.six}) is
\begin{equation}
  {\hbar\over2}|\partial_x\theta(x)|\ll|p(x)|,
\label{eq:two.eleven}
\end{equation}
which is expected to be satisfied when the coupling constant~$g$ is
small and the incident particle is sufficiently energetic.

%%%%%%%%%%%%%%%%%%%%%%%%%%%%%%%%%%%%%%%%%%%%%%%%%%%%%%%%%%%%%%%%%%%%
\section{Landau-Zener formula}
\label{sec:three}

As an illustration of the duality discussed in Section~\ref{sec:two},
we re-examine a perturbative derivation of the Landau-Zener formula
in both of the adiabatic and diabatic pictures~\cite{1,5,8}. For
definiteness, we shall assume $V_1'(0)>V_2'(0)$ as in~Fig.~1.

Let us start with the adiabatic picture with weak potential crossing
interaction. Since the gauge field generally vanishes,
$\partial_x\theta(x)\rightarrow0$ for $|x|\rightarrow\infty$, we can
define the {\it asymptotic\/} states in terms of the eigenstates
of~$H_0'$~(\ref{eq:two.seven}). We define the initial and final
states $\Phi_i$ and~$\Phi_f$ by
\begin{equation}
   \Phi_i(x)=\pmatrix{\varphi_1(x)\cr 0\cr},\quad
   \Phi_f(x)=\pmatrix{0\cr \varphi_2(x)\cr},
\label{eq:three.one}
\end{equation}
which satisfy
\begin{eqnarray}
   \left[{1\over2m}\hat p^2+U_1(x)\right]\varphi_1(x)
   &=&E\varphi_1(x),
\nonumber\\
   \left[{1\over2m}\hat p^2+U_2(x)\right]\varphi_2(x)
   &=&E\varphi_2(x).
\label{eq:three.two}
\end{eqnarray}
We then obtain the potential curve crossing probability due to the
perturbation~$H_I'$~(\ref{eq:two.eight})
\begin{equation}
   w(i\rightarrow f)
   ={2\pi\over\hbar}|\langle\Phi_f|H_I'|\Phi_i\rangle|^2.
\label{eq:three.three}
\end{equation}
The transition matrix element is given by
\begin{eqnarray}
   \langle\Phi_f|H_I'|\Phi_i\rangle&=&
   -{\hbar^2\over4m}\int_{-\infty}^\infty dx\,\partial_x\theta(x)
\\
  &\times&\left[-\varphi_2'(x)\varphi_1(x)
                +\varphi_2(x)\varphi_1'(x)\right].
\nonumber
\label{eq:three.four}
\end{eqnarray}

To evaluate the matrix element, we use the WKB wave
functions~\cite{4}:
\begin{equation}
   \varphi_1(x)=\cases{
   \displaystyle
   {C_1\over2\sqrt{|p_1(x)|}}
   \exp\left[-{1\over\hbar}\int_{a_1}^xdx\,|p_1(x)|\right],\cr
   \displaystyle
   {C_1\over\sqrt{p_1(x)}}
   \cos\left[{1\over\hbar}\int_x^{a_1}dx\,p_1(x)
   -{\pi\over4}\right]\cr}
\label{eq:three.five}
\end{equation}
for $x>a_1$ and~$x<a_1$, respectively, and
\begin{equation}
   \varphi_2(x)=\cases{
   \displaystyle
   {C_2\over2\sqrt{|p_2(x)|}}
   \exp\left[-{1\over\hbar}\int_{a_2}^xdx\,|p_2(x)|\right],\cr
   \displaystyle
   {C_2\over\sqrt{p_2(x)}}
   \cos\left[{1\over\hbar}\int_x^{a_2}dx\,p_2(x)
   -{\pi\over4}\right]\cr}
\label{eq:three.six}
\end{equation}
for $x>a_2$ and~$x<a_2$, respectively. The semi-classical momenta in
the adiabatic picture are defined by
\begin{equation}
   p_{1,2}(x)\equiv\sqrt{2m[E-U_{1,2}(x)]},
\label{eq:three.seven}
\end{equation}
and $a_1$ and~$a_2$ denote the classical turning points (see~Fig.~2).
The normalization of~$\varphi_1(x)$ is chosen as~$C_1=2\sqrt{m}$ to
make the probability flux of the incident wave unity. On the other
hand, the final state wave function in~(\ref{eq:three.three}) has to
be normalized by the delta function with respect to the energy,
$\langle\Phi_2'|\Phi_2\rangle=\delta(E_2'-E_2)$ and this specifies
$C_2=2\sqrt{m}/\sqrt{2\pi\hbar}$.

We estimate the matrix element~(15) by using the oscillating parts
of the wave functions (\ref{eq:three.five}) and~(\ref{eq:three.six}).
This treatment is justified if the following conditions are
satisfied:\\
(i)~$|p(0)|\rightarrow{\rm large}$ and $m\rightarrow{\rm large}$ with
$v=|p(0)|/m$ kept fixed such that non-relativistic treatment is valid
in the physically relevant region.\\
(ii)~$g\rightarrow{\rm small}$, but with
\begin{equation}
   {1\over g}\ll{1\over2m}|p(0)|^2
\label{eq:three.eleven}
\end{equation}
to ensure~(\ref{eq:two.eleven}) and the condition
\begin{equation}
   \beta\ll a,
\label{eq:three.twelve}
\end{equation}
where $a$~is an average turning point and $\beta$~is a typical
geometrical extension of~$\partial_x\theta(x)$. If
(\ref{eq:three.twelve})~is satisfied, we can estimate the matrix
element by using the oscillating parts of wave functions only since
$\partial_{x}\theta(x)$~rapidly goes to zero for~$|x|\gg\beta$ on the
real axis.

The integral~(15) is then written as
\begin{eqnarray}
   &&\langle\Phi_f|H_I'|\Phi_i\rangle
   \simeq-{i\hbar C_1C_2\over 8m}\int dx\,\partial_x\theta(x)
\\
   &&\times\biggl\{
   \exp\left[{i\over\hbar}\int_{a_1}^xdx\,p_1(x)
             -{i\over\hbar}\int_{a_2}^xdx\,p_2(x)\right]
\nonumber\\
   &&-\exp\left[-{i\over\hbar}\int_{a_1}^xdx\,p_1(x)
             +{i\over\hbar}\int_{a_2}^xdx\,p_2(x)\right]\biggr\},
\nonumber
\label{eq:three.nine}
\end{eqnarray}
where we have set $p_1(x)/p_2(x)=1$ in the prefactors. This is
justified if~$\hbar/|p(0)|\ll\beta$, the characteristic length scale
of the present problem, by letting~$p(0)$ large as is specified
in~(i). Therefore we need to evaluate an integral of the form
\begin{eqnarray}
   I&\equiv&\int_{-\infty}^\infty dx\,\partial_x\theta(x)
\\
   &\times&\exp\left[{i\over\hbar}\int_{a_1}^xdx\,p_1(x)
             -{i\over\hbar}\int_{a_2}^xdx\,p_2(x)\right].
\nonumber
\label{eq:three.ten}
\end{eqnarray}

We here present an explicit evaluation of~(22) for the linear
potential crossing problem, $V_1(x)=V_1'(0)x$ and~$V_2(x)=V_2'(0)x$,
on the basis of local data without referring to Stokes phenomena. For
sufficiently large energy,
$E-(U_1+U_2)/2\gg(U_1-U_2)/2$, the difference of momenta can be
approximated as [see~(\ref{eq:two.nine})],
\begin{eqnarray}
   &&\int_0^xdx\,[p_1(x)-p_2(x)]
\\
   &&\simeq-\int_0^xdx\,{2\over v(x)g\beta}\,\sqrt{x^2+\beta^2}
\nonumber
\label{eq:three.thirteen}
\end{eqnarray}
where we used 
\begin{eqnarray}
   f(x)&=&g{V_1'(0)-V_2'(0)\over2}x\equiv{x\over\beta},
\\
   v(x)&\equiv&{1\over m}
   \sqrt{2m\left[E-{U_1(x)+U_2(x)\over2}\right]}
\nonumber
\label{eq:three.fourteen}
\end{eqnarray}
and $v(x)$~is approximated to be a constant $v=v(0)$ in the
following. We also have from~(\ref{eq:two.ten})
\begin{equation}
   \partial_x\theta(x)=-{\beta\over x^2+\beta^2}
\label{eq:three.fifteen}
\end{equation}
and thus
\begin{eqnarray}
   &&I\simeq
   -\exp\left[{i\over\hbar}\int_{a_1}^0dx\,p_1(x)
         -{i\over\hbar}\int_{a_2}^0dx\,p_2(x)\right]
\nonumber\\
   &&\times\int_{-\infty}^\infty dx\,{\beta\over x^2+\beta^2}
   \exp\left(-{2i\over\hbar vg\beta}
   \int_0^xdx\,\sqrt{x^2+\beta^2}\right)
\nonumber\\
   &&=-\exp\left[{i\over\hbar}\int_{a_1}^0dx\,p_1(x)
      -{i\over\hbar}\int_{a_2}^0dx\,p_2(x)\right]
\nonumber\\
   &&\times\int_{-\infty}^\infty dx\,\exp[-i\alpha F(x)]
\label{eq:three.sixteen}
\end{eqnarray}
where
\begin{eqnarray}
  F(x)&\equiv&\int_0^xdx\,\sqrt{x^2+1}+{1\over i\alpha}\ln(x^2+1),
\nonumber\\
  \alpha&\equiv&{2\beta\over\hbar vg}
   ={4\over\hbar vg^2[V_1'(0)-V_2'(0)]}>0. 
\label{eq:three.seventeen}
\end{eqnarray}
We evaluate the integral~(\ref{eq:three.sixteen}) by a saddle point
approximation with respect to~$\alpha$. We thus seek the saddle point
\begin{equation}
   F'(x)=\sqrt{x^2+1}+{1\over i\alpha}{2x\over x^2+1}=0,
\label{eq:three.eighteen}
\end{equation}
which is located between the real axis and the pole
positions~$x=\pm i$ of~$\partial_x\theta(\beta x)$ so that we can
smoothly deform the integration contour; these poles also coincide
with the complex potential crossing points. If one sets~$x=iy$
in~(\ref{eq:three.eighteen}) for $-1<y<1$, one has
\begin{equation}
   \sqrt{1-y^2}=-{1\over\alpha}{2y\over1-y^2}
\label{eq:three.nineteen}
\end{equation}
which has a {\it unique\/} solution
\begin{equation}
   x_s=iy_s\simeq-i+{i\over2}\left({2\over\alpha}\right)^{2/3}
\label{eq:three.twenty}
\end{equation}
for {\it large\/}~$\alpha$. (The complex conjugate of~$x_s$ is
located in the second Riemann sheet.) For this value of the saddle
point
\begin{eqnarray}
   F(x_s)&=&\int_0^{x_s}dx\,\sqrt{x^2+1}
   +{1\over i\alpha}\ln\left({2\over\alpha}\right)^{2/3}
\nonumber\\
   &\simeq&-{\pi i\over4}+{2\over3}{i\over\alpha}
   +{1\over i\alpha}\ln\left({2\over\alpha}\right)^{2/3},
\nonumber\\
   F''(x_s)&\simeq&-3i\left({\alpha\over2}\right)^{1/3}.
\label{eq:three.twentyone}
\end{eqnarray}
We thus have a Gaussian integral which decreases in the direction
{\it parallel\/} to the real axis
\begin{eqnarray}
   I&\simeq&-\left({\alpha\over2}\right)^{2/3}
            e^{2/3}\exp\left(-{\pi\alpha\over4}\right)
\\
   &\times&\exp\left[{i\over\hbar}\int_{a_1}^0dx\,p_1(x)
             -{i\over\hbar}\int_{a_2}^0dx\,p_2(x)\right]
\nonumber\\
   &\times&
   \int_{-\infty}^\infty dx\,
   \exp\left[-3\left(\alpha\over2\right)^{4/3}(x-x_s)^2\right]
\nonumber\\
   &=&
   -\sqrt{\pi\over3}e^{2/3}\exp\left(-{\pi\alpha\over4}\right)
\nonumber\\
   &\times&\exp\left[{i\over\hbar}\int_{a_1}^0dx\,p_1(x)
         -{i\over\hbar}\int_{a_2}^0dx\,p_2(x)\right]
\nonumber 
\label{eq:three.twentytwo}
\end{eqnarray}
{}From~(\ref{eq:three.nine}) we obtain
\begin{eqnarray}
   &&\langle\Phi_f|H_I'|\Phi_i\rangle\simeq
   -\sqrt{\pi\over3}e^{2/3}\sqrt{\hbar\over2\pi}
\nonumber\\
   &&\times\sin\left\{{1\over\hbar}
   \left[\int_{a_1}^{0}dx\,p_1(x)-\int_{a_2}^{0}dx\,p_2(x)\right]
   \right\}
\nonumber\\
   &&\times
   \exp\left\{-{\pi\over\hbar vg^2[V_1'(0)-V_2'(0)]}
   \right\}.
\label{eq:three.twentythree}
\end{eqnarray}
It is interesting that the numerical value of the coefficient of the
above expression, $\sqrt{\pi}e^{2/3}/\sqrt{3}=1.99317$, is very close
to the canonical value~$2$~\cite{4}, and we replace it by~$2$ in the
following. As for the past analysis of the prefactor in the
time-dependent perturbation theory, see papers in~\cite{13}. We thus
have the transition probability from~(\ref{eq:three.three})
\begin{eqnarray}
   &&w(i\rightarrow f)
\nonumber\\
   &&\simeq
   4\sin^2\left\{{1\over\hbar}\,
   \left[\int_{a_1}^0dx\,p_1(x)-\int_{a_2}^0dx\,p_2(x)\right]\right\}
\nonumber\\
    &&\times
   \exp\left\{-{2\pi\over\hbar vg^2[V_1'(0)-V_2'(0)]}\right\}
\nonumber\\
   &&\simeq2\exp\left\{-{2\pi\over\hbar vg^2[V_1'(0)-V_2'(0)]}
     \right\}
\label{eq:three.twentyfour}
\end{eqnarray}
where we replaced the square of sine function by its average~$1/2$
in the final expression. We emphasize that the numerical coefficient
of~$w(i\rightarrow f)$ is fixed by time-independent perturbation
theory and the local data without referring to global Stokes
phenomena; this is satisfactory since linear potential crossing is a
locally valid idealization.

We interpret that $w(i\rightarrow f)$~in~(\ref{eq:three.twentyfour})
expresses {\it twice\/} the non-adiabatic transition probability.
Notice that our initial state wave function contains the reflection
wave as well as the incident wave. Therefore the transition
probability per {\it one\/} crossing is given by the half
of~(\ref{eq:three.twentyfour}),
\begin{equation}
   P(1\rightarrow2)\simeq
   \exp\left\{-{2\pi\over\hbar vg^2[V_1'(0)-V_2'(0)]}\right\},
\label{eq:three.twentyseven}
\end{equation}
which is the celebrated Landau-Zener formula~\cite{4,5}. Our
perturbative derivation presented here is conceptually much simpler
than the past works~\cite{1,4,5,8}, and it should be useful for a
pedagogical purpose also.

It is interesting to study the same problem in the diabatic picture
in~Fig.~1 with~$H_I=\sigma_1/g$ for large~$g$. The evaluation of the
matrix element is the standard one described in the textbook of
Landau and Lifshitz~\cite{4}, for example. We have for $E>0$,
\begin{eqnarray}
   &&w(i\rightarrow f)\simeq
   {8\pi\over\hbar v(0)g^2[V_1'(0)-V_2'(0)]}
\nonumber\\
   &&\times\cos^2\left[{1\over\hbar}\int_0^{a_2}dx\,p_2(x)
               -{1\over\hbar}\int_0^{a_1}dx\,p_1(x)
               -{\pi\over4}\right]
\nonumber\\
   &&\simeq{4\pi\over\hbar vg^2[V_1'(0)-V_2'(0)]}.
\label{eq:three.thirtyfour}
\end{eqnarray}
[$v(0)$ is the velocity at the crossing point, $v(0)=\sqrt{2E/m}$.]
We again interpret~(\ref{eq:three.thirtyfour}) as twice the potential
crossing probability because our initial state wave function contains
the reflection wave as well as the incident wave. The transition
probability per one potential crossing is given by the half
of~(\ref{eq:three.thirtyfour}).

A simple {\it interpolating\/} formula, which
reproduces~(\ref{eq:three.twentyfour}) in the weak coupling limit and
(\ref{eq:three.thirtyfour}) in the strong coupling limit, is given by
\begin{eqnarray}
  &&w(i\rightarrow f)\simeq
   2\exp\left\{-{2\pi\over\hbar vg^2[V_1'(0)-V_2'(0)]}\right\}
\nonumber\\
  &&\times\left(1-
    \exp\left\{-{2\pi\over\hbar vg^2[V_1'(0)-V_2'(0)]}\right\}
   \right)
\label{eq:three.thirtysix}
\end{eqnarray}
This expression is also consistent with the (semi-classical)
conservation of probability~\cite{4}.

Motivated by duality, we re-examined a perturbative derivation of
the Landau-Zener formula, and we re-derived the
formula~(\ref{eq:three.twentyfour}) including its numerical
coefficient on the basis of perturbation theory. However, our final
result~(\ref{eq:three.twentyfour}) in the adiabatic picture does not
contain the coupling constant as a prefactor. This is related to an
interesting topological object in the present formulation. From the
definition of~(\ref{eq:two.ten}), the ``gauge field'' satisfies the
relation
\begin{eqnarray}
   \int_{-\infty}^\infty dx\,\partial_x\theta(x)
   &=&\theta(\infty)-\theta(-\infty)
\nonumber\\
   &=&-\pi,
\label{eq:three.thirtyseven}
\end{eqnarray}
which is {\it independent\/} of the coupling constant~$g$; we assume 
$f(x)\rightarrow\pm\infty$ for~$x\rightarrow\pm\infty$, respectively.
Because of this kink-like topological behavior of~$\theta(x)$, the
coupling constant does not appear as a prefactor of the matrix
element in perturbation theory if the wave functions spread over the
range which well covers the geometrical size
of~$\partial_x\theta(x)$. The precise criterion of the validity of
perturbation theory is thus given by~(\ref{eq:two.eleven}): This
condition is in fact satisfied if the conditions
(\ref{eq:three.eleven})--(\ref{eq:three.twelve}) are satisfied. For
small values of~$x$, the small coupling~$g$ helps to
satisfy~(\ref{eq:two.eleven}). Even for the values of~$x$ near the
average turning point~$a$, we have
\begin{equation}
   {\hbar\over2}|\partial_x\theta(a)|
   \simeq{1\over2}\left({\beta\over a}\right){\hbar\over a}\ll
   {\hbar\over a}\simeq|p(a)|,
\label{eq:three.thirtyeight}
\end{equation}
where $\beta$~stands for the typical geometrical size
of~$\partial_x\theta(x)$. The estimate in the left hand side is based
on linear potentials~(\ref{eq:three.fifteen}), but we expect that the
condition is satisfied for more general potentials as well. We thus
clarified the basic mechanism why the prefactor of the Landau-Zener
formula~(\ref{eq:three.twentyseven}) should come out to be very close
to unity in time-independent perturbation theory.

%%%%%%%%%%%%%%%%%%%%%%%%%%%%%%%%%%%%%%%%%%%%%%%%%%%%%%%%%%%%%%%%%%%%
\section{Discussions}
\label{sec:four}

Motivated by the presence of interesting weak and strong duality in
the model Hamiltonian~(\ref{eq:two.one}) of potential curve crossing,
we re-examined a perturbative approach to potential crossing
phenomena. We have shown that straightforward
{\it time-independent\/} perturbation theory combined with the zeroth
order WKB wave functions provides a reliable description of general
potential crossing phenomena. Our analysis is based on the local data
as much as possible without referring to global Stokes phenomena.
Formulated in this manner, perturbation theory becomes more flexible
to cover a wide range of problems.

The effects of dissipative interactions on macroscopic quantum
tunneling have been extensively analyzed in the path integral
formalism~\cite{14} and also in the canonical (field theoretical)
formalism~\cite{15}. It is generally accepted that the Ohmic
dissipation suppresses the macroscopic quantum coherence; in fact,
an attractive idea of a dissipative phase transition has been
suggested~\cite{14}.

It is plausible that the effects of potential curve crossing with
nearby potentials influence the quantum coherence of the two
degenerate ground states. One can in fact confirm that the potential
curve crossing generally suppress the quantum coherence by using the
perturbation theory for both limits of strong and weak curve crossing
interactions~\cite{11}. From a view point of symmetry, the lowest
order perturbation in the present problem and the dissipative
interaction in the Caldeira-Legget model~\cite{14} both correspond to
a dipole approximation. However, a perturbative analysis of basically
non-perturbative tunneling phenomena requires a great care.
In~Ref.~\cite{11}, an explicit calculation of rather limited
configurations has been performed, which in fact indicates the
general suppression of quantum coherence by potential curve crossing.
This suppression phenomenon of quantum coherence may become important
in the future when one takes the effects of the environment into
account in the analysis of potential curve crossing processes.

{}From a view point of general gauge theory, it is not unlikely that
the electric-magnetic duality in conventional gauge theory is also
related to some generalized form of potential crossing in the
so-called moduli space~\cite{12}. We hope that our work may turn out
to be relevant from this view point also.

%%%%%%%%%%%%%%%%%%%%%%%%%%%%%%%%%%%%%%%%%%%%%%%%%%%%%%%%%%%%%%%%%%%%

%%%%%%%%%%%%%%%%%%%%%%%%%%%%%%%%%%%%%%%%%%%%%%%%%%%%%%%%%%%%%%%%%%%%
\end{document}